\documentclass[letterpaper, 10 pt, journal, twoside]{IEEEtran}
\pdfoutput=1
\usepackage{hyperref}
\hypersetup{hidelinks,pdfnewwindow,breaklinks} 
\usepackage{graphicx}
\usepackage{mathptmx} 
\usepackage{times} 
\usepackage{cite}
\usepackage{amsmath,amssymb,amsfonts}
\usepackage[ruled,vlined]{algorithm2e}
\usepackage{textcomp}
\usepackage{bm}
\usepackage{listings}
\usepackage[caption=false,font=normalsize,labelfon
t=sf,textfont=sf]{subfig}
\usepackage{mathtools}
\usepackage{color}
\usepackage{import}
\usepackage{boxedminipage}

\definecolor{caltechblue}{rgb}{0,0.2314,0.2980}
\definecolor{caltechgreen}{rgb}{0,0.3451,0.3137}
\definecolor{caltechorange}{rgb}{1,0.4235,0.0406}

\makeatletter
\def\ps@IEEEtitlepagestyle{
\def\@oddhead{\hbox{}\@IEEEheaderstyle\leftmark\hfil\thepage}\relax
\def\@evenhead{\@IEEEheaderstyle\thepage\hfil\leftmark\hbox{}}\relax
  \def\@oddfoot{\mycopyrightnotice}
  \def\@evenfoot{}
}
\def\mycopyrightnotice{
  {\footnotesize
  \begin{boxedminipage}{\textwidth}
  \centering
  © 2020 IEEE. Personal use of this material is permitted. Permission from IEEE must be obtained for all other uses, in any current or future media, including reprinting/republishing this material for advertising or promotional purposes, creating new collective works, for resale or redistribution to servers or lists, or reuse of any copyrighted component of this work in other works. Digital Object Identifier (DOI): {\color{caltechgreen}\underline{\href{https://ieeexplore.ieee.org/document/9115010}{10.1109/LCSYS.2020.3001646}}}
  \end{boxedminipage}
  }
}

\def\BibTeX{{\rm B\kern-.05em{\sc i\kern-.025em b}\kern-.08em
    T\kern-.1667em\lower.7ex\hbox{E}\kern-.125emX}}
\newtheorem{theorem}{Theorem}

\newtheorem{lemma}{Lemma}

\newtheorem{proposition}{Proposition}
 
\newtheorem{remark}{Remark}

\newtheorem{corollary}{Corollary}

\newcommand{\ie}{{i}.{e}.}

\newcommand{\st}{{s}.{t}.}

\DeclareMathOperator{\sym}{sym}

\title{Neural Contraction Metrics for Robust Estimation and Control: A Convex Optimization Approach
}

\author{Hiroyasu Tsukamoto, \IEEEmembership{Member, IEEE}, and Soon-Jo Chung, \IEEEmembership{Senior Member, IEEE}
\thanks{This work was in part funded by the Jet Propulsion Laboratory, California Institute of Technology and the Raytheon Company.}
\thanks{The authors are with the Graduate Aerospace Laboratories, California Institute of Technology, Pasadena, CA, USA. E-mail: {\tt\small \{htsukamoto, sjchung\}@caltech.edu}), Code: {\color{caltechgreen}\underline{\url{https://github.com/astrohiro/ncm}}}}.
}

\pdfoutput=1 
\begin{document}
\maketitle
\begin{abstract}
This paper presents a new deep learning-based framework for robust nonlinear estimation and control using the concept of a \underline{N}eural \underline{C}ontraction \underline{M}etric (NCM). The NCM uses a deep long short-term memory recurrent neural network for a global approximation of an optimal contraction metric, the existence of which is a necessary and sufficient condition for exponential stability of nonlinear systems. The optimality stems from the fact that the contraction metrics sampled offline are the solutions of a convex optimization problem to minimize an upper bound of the steady-state Euclidean distance between perturbed and unperturbed system trajectories. We demonstrate how to exploit NCMs to design an online optimal estimator and controller for nonlinear systems with bounded disturbances utilizing their duality. The performance of our framework is illustrated through Lorenz oscillator state estimation and spacecraft optimal motion planning problems.
\end{abstract}
\section{Introduction}
\label{introduction}
\IEEEPARstart{P}{rovably} stable and optimal state estimation and control algorithms for a class of nonlinear dynamical systems with external disturbances are essential to develop autonomous robotic explorers operating remotely on land, in water, and in deep space. In these next generation missions, these robots are supposed to intelligently perform complex tasks with their limited computational resources, which are not necessarily powerful enough to run optimization algorithms in real-time.

\begin{figure}
    \centering
    \setlength\belowcaptionskip{-0.6\baselineskip}
    \includegraphics[width=85mm]{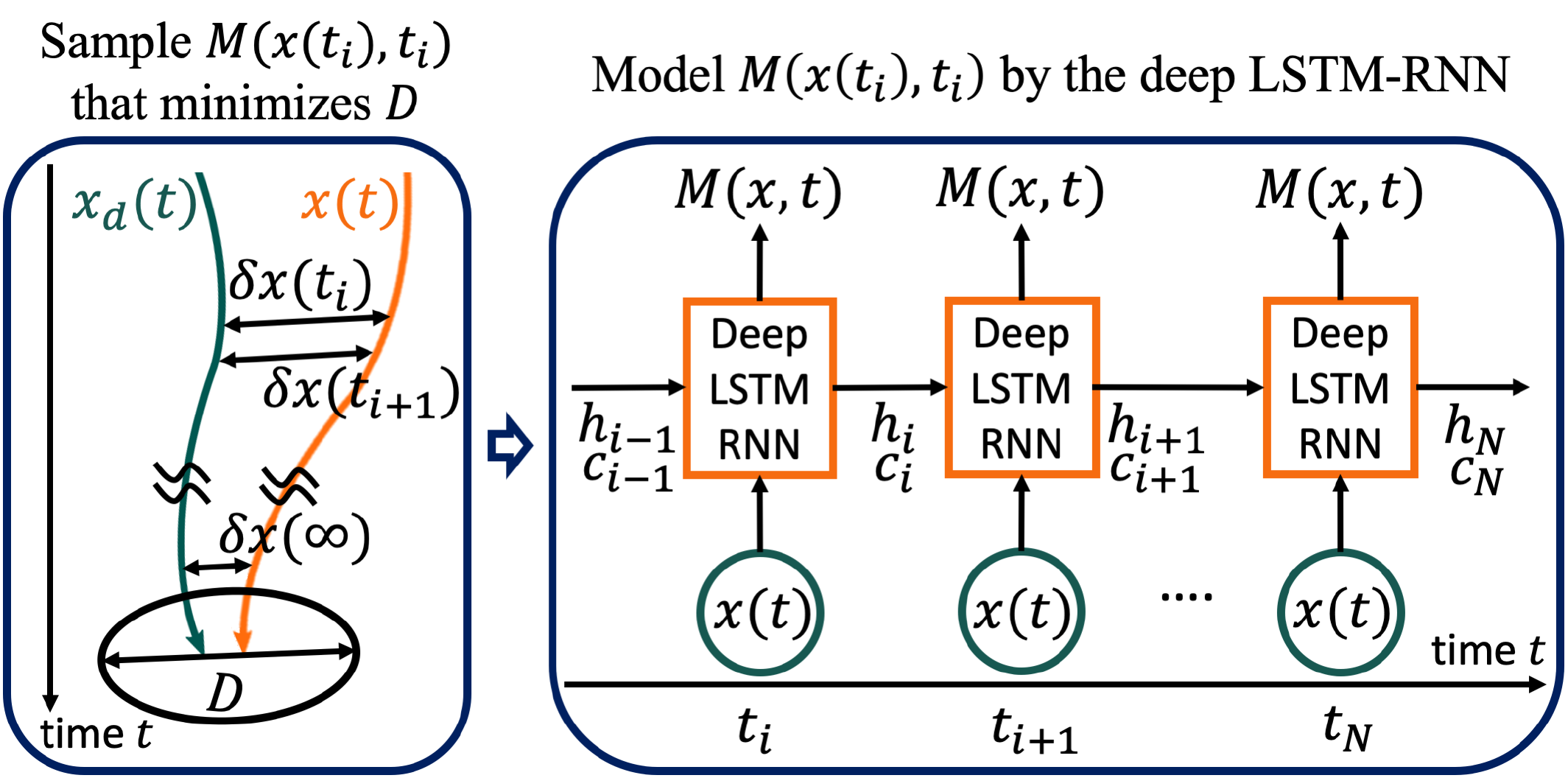}
    \caption{Illustration of the NCM: $M(x,t)$ denotes the optimal contraction metric; $x(t)$ and $x_d(t)$ denote perturbed and unperturbed system trajectories; $h_i$ and $c_i$ denote the hidden states of the deep LSTM-RNN, respectively.}
    \label{ncmdrawing}
\end{figure}

Our main contribution is to introduce a Neural Contraction Metric (NCM), a global representation of optimal contraction metrics sampled offline by using a deep Long Short-Term Memory Recurrent Neural Network (LSTM-RNN) (see Fig.~\ref{ncmdrawing}), and thereby propose a new framework for provably stable and optimal online estimation and control of nonlinear systems with bounded disturbances, which only requires one function evaluation at each time step. A deep LSTM-RNN~\cite{doi:10.1162/neco.1997.9.8.1735,6638947} is a recurrent neural network with an improved memory structure proposed to circumvent gradient vanishing~\cite{279181} and is a universal approximator of continuous curves~\cite{FUNAHASHI1993801}. Contrary to previous works, the convex optimization-based sampling methodology in our framework allows us to obtain a large enough dataset of the optimal contraction metric without assuming any hypothesis function space. These sampled metrics, the existence of which is a necessary and sufficient condition for exponential convergence \cite{contraction}, can be approximated with arbitrary accuracy due to the high representational power of the deep LSTM-RNN.
We remark that this approach can be used with learned dynamics~\cite{8794351} as a nominal model is assumed to be given. Also, this is distinct from Lyapunov neural networks designed to estimate a largest safe region for deterministic systems \cite{spencer18lyapunovnn,NIPS2019_8587}: the NCM provides provably stable estimation and control policies, which have a duality in their differential dynamics and are optimal in terms of disturbance attenuation. The NCM construction is summarized as follows.

In the offline phase, we sample contraction metrics by solving an optimization problem with exponential stability constraints, the objective of which is to minimize an upper bound of the steady-state Euclidean distance between perturbed and unperturbed system trajectories. In this paper, we present a convex optimization problem equivalent to this problem, thereby exploiting the differential nature of contraction analysis that enables Linear Time-Varying (LTV) systems-type approaches to Lyapunov function construction. For the sake of practical use, the sampling methodology is reduced to a much simpler formulation than those of \cite{observer, mypaper, mypaperTAC} derived for It\^{o} stochastic nonlinear systems. These optimal contraction metrics are sampled using the computationally efficient numerical methods for convex programming~\cite{citeulike:163662,cvxpy,mosek} and then modeled by the deep LSTM-RNN as depicted in Fig.~\ref{ncmdrawing}. In the online phase, contraction metrics at each time instant are computed by the NCM to obtain the optimal feedback estimation and control gain or a bounded error tube for robust motion planning~\cite{chung2013phase,7989693}.

We illustrate how to design an optimal NCM-based estimator and controller for nonlinear systems with bounded disturbances, utilizing the estimation and control duality in differential dynamics analogous to the one of the Kalman filter and Linear Quadratic Regulator (LQR) in LTV systems. Their performance is demonstrated using Lorenz oscillator state estimation and spacecraft optimal motion planning problems.
\subsubsection*{Related Work}
Contraction analysis, as well as Lyapunov theory, is one of the most powerful tools in analyzing the stability of nonlinear systems~\cite{contraction}. It studies the differential (virtual) dynamics for the sake of incremental stability by means of a contraction metric, the existence of which leads to a necessary and sufficient characterization of exponential stability of nonlinear systems. Finding an optimal contraction metric for general nonlinear systems is, however, almost as difficult as finding an optimal Lyapunov function.

Several numerical methods have been developed for finding contraction metrics in a given hypothesis function space. A natural application of this concept is to represent their candidate as a linear combination of some basis functions~\cite{664157,JOHANSEN20001617,LyapunovRBF,1184414}. In ~\cite{AYLWARD20082163,ccm}, a tractable framework to construct contraction metrics for dynamics with polynomial vector fields is proposed by relaxing the stability conditions to the sum of squares conditions. Although it is ideal to use a larger number of basis functions seeking for a more optimal solution, the downside of this approach is that the problem size grows exponentially with the number of variables and basis functions \cite{sos_dissertation}.

We could thus alternatively rely on numerical schemes for sampling data points of a Lyapunov function without assuming any hypothesis function space. This includes the state-dependent Riccati equation method~\cite{sddre,CIMEN20083761} and it is proposed in~\cite{observer,mypaper,mypaperTAC} that this framework can be improved to obtain an optimal contraction metric, which minimizes an upper bound of the steady-state mean squared tracking error for nonlinear stochastic systems. However, solving a nonlinear system of equations or an optimization problem at each time instant is not suitable for systems with limited online computational capacity. The NCM addresses this issue by approximating the sampled solutions by the LSTM-RNN.
\section{Preliminaries}
\label{preliminaries}
We use the notation $\|x\|$ and $\|A\|$ for the Euclidean and induced 2-norm, and $A \succ 0$, $A \succeq 0$, $A \prec 0$, and $A \preceq 0$ for positive definite, positive semi-definite, negative definite, and negative semi-definite matrices, respectively. Also, $\sym(A) = A+A^T$ and $I$ denotes the identity matrix.
We first introduce some preliminaries that will be used to construct an NCM.
\subsection{Contraction Analysis for Incremental Stability}
Consider the following perturbed nonlinear system:
\begin{align}
    \label{nonlin_per}
    \dot{x}(t) = f(x(t),t)+d(t)
\end{align}
where $t\in \mathbb{R}_{\geq0}$, $x:\mathbb{R}_{\geq0} \to \mathbb{R}^{n}$, $f:\mathbb{R}^n\times\mathbb{R}_{\geq0} \to \mathbb{R}^{n}$, and $d:\mathbb{R}_{\geq0} \to \mathbb{R}^{n}$ with $\overline{d}=\sup_{t \geq 0}\|d(t)\| < +\infty$.
\begin{theorem}
\label{RobustCont}
Let $x_1(t)$ and $x_2(t)$ be the solution of (\ref{nonlin_per}) with $d(t) = 0$ and $d(t) \neq 0$, respectively. Suppose there exist $M(x,t) = \Theta (x,t)^T\Theta(x,t) \succ 0$, $\alpha > 0$, and $0 < \underline{\omega},\overline{\omega} < \infty$ \st{}
\begin{align}
\label{deterministic_contraction}
&\dot{M}(x,t)+\sym{\left(M(x,t)\frac{\partial f(x,t)}{\partial x}\right)} \preceq -2\alpha M(x,t),~\forall x,t \\
\label{Mcon}
& \left(\frac{1}{\overline{\omega}}\right)I \preceq M(x,t) \preceq \left(\frac{1}{\underline{\omega}}\right)I,~\forall x,t.
\end{align}
Then the smallest path integral is exponentially bounded, thereby yielding a bounded tube of states:
\begin{align}
    \label{dist_dist}
    \int^{x_2}_{x_1}\|\delta x\|-{R(0)}{\sqrt{\overline{\omega}}}e^{-\alpha t} \leq \frac{\overline{d}}{\alpha}\sqrt{\frac{\overline{\omega}}{\underline{\omega}}} \leq \frac{\overline{d}\overline{\omega}}{\alpha\underline{\omega}}
\end{align}
where $R(t) = \int_{x_1}^{x_2}\|\Theta(x,t) \delta x(t)\|, \forall t$.
\end{theorem}
\begin{IEEEproof}
Differentiating $R(t)$ yields $\dot{R}(t)+\alpha R(t) \leq \|\Theta(x(t),t)d(t)\|$ (see~\cite{contraction}). Since we have $\|\Theta(x(t),t)d(t)\| \leq \overline{d}/\sqrt{\underline{\omega}}$, applying the comparison lemma~\cite{Khalil:1173048} results in $R(t) \leq R(0)e^{-\alpha t}+\overline{d}/(\alpha\sqrt{\underline{\omega}})$. Rewriting this inequality using the relations $\sqrt{\overline{\omega}} R(t) \geq \int^{x_2}_{x_1}\|\delta x\|$ and  $1 \leq \sqrt{\overline{\omega}/\underline{\omega}} \leq \overline{\omega}/\underline{\omega}$ due to $0 < \underline{\omega} \leq \overline{\omega}$ completes the proof. Input-to-state stability and finite-gain $\mathcal{L}_p$ stability follow from (\ref{dist_dist}) (see~\cite{chung2013phase}).
\end{IEEEproof}
\subsection{Deep LSTM-RNN}
An LSTM-RNN is a neural network designed for processing sequential data with inputs $\{x_i\}_{i=0}^N$ and outputs $\{y_i\}_{i=0}^N$, and defined as $y_i = W_{hy}h_i+b_y$ where $h_i = \phi(x_i,h_{i-1},c_{i-1})$.
The activation function $\phi$ is given by the following relations: $h_{i} = o_{i}\tanh(c_{i})$, $o_{i} = \sigma(W_{xo}x_{i}+W_{ho}h_{i-1}+W_{co}c_{i}+b_o)$, $c_{i} = f_{i}c_{i-1}+\iota_{i}\tanh(W_{xc}x_{i}+W_{hc}h_{i-1}+b_c)$, $f_{i} = \sigma(W_{xf}x_{i}+W_{hf}h_{i-1}+W_{cf}c_{i-1}+b_f)$, and $\iota_{i} = \sigma(W_{xi}x_{i}+W_{hi}h_{i-1}+W_{ci}c_{i-1}+b_i)$, where $\sigma$ is the logistic sigmoid function and $W$ and $b$ terms represent weight matrices and bias vectors to be optimized, respectively. The deep LSTM-RNN can be constructed by stacking multiple of these layers~\cite{doi:10.1162/neco.1997.9.8.1735,6638947}.

Since contraction analysis for discrete-time systems leads to similar results~\cite{contraction,mypaperTAC}, we define the inputs $x_i$ as discretized states $\{x_i = x(t_i)\}_{t=0}^N$, and the outputs $y_i$ as non-zero components of the unique Cholesky decomposition of the optimal contraction metric, as will be discussed in Sec.~\ref{ncmsec}.
\section{Neural Contraction Metric (NCM)}
\label{ncmsec}
This section presents an algorithm to obtain an NCM depicted in Fig. \ref{ncmdrawing}.
\subsection{Convex Optimization-based Sampling of Contraction Metrics (CV-STEM)}
We derive one approach to sample contraction metrics by using the \underline{C}on\underline{V}ex optimization-based \underline{S}teady-state \underline{T}racking \underline{E}rror \underline{M}inimization (CV-STEM) method~\cite{mypaper,mypaperTAC}, which could handle the control design of It\^{o} stochastic nonlinear systems. In this section, we propose its simpler formulation for nonlinear systems with bounded disturbances in order to be of practical use in engineering applications.

By Theorem~\ref{RobustCont}, the problem to minimize an upper bound of the steady-state Euclidean distance between the trajectory $x_1(t)$ of the unperturbed system and $x_2(t)$ of the perturbed system (\ref{nonlin_per}) ((\ref{dist_dist}) as $t\to \infty$) can be formulated as follows:
\begin{align}
\label{nonlin_opt_ccm}
{J}_{NL}^* = \min_{\underline{\omega}>0,\overline{\omega}>0,W \succ 0} \frac{\overline{d}\overline{\omega}}{\alpha\underline{\omega}}\text{  \st{} }\text{(\ref{deterministic_contraction}) and (\ref{Mcon})}
\end{align}
where $W(x,t) = M(x,t)^{-1}$ is used as a decision variable instead of $M(x,t)$. We assume that the contraction rate $\alpha$ and disturbance bound $\overline{d}$ are given in (\ref{nonlin_opt_ccm}) (see Remark~\ref{howtochosealpha} on how to select $\alpha$).
We need the following lemma to convexify this nonlinear optimization problem.
\begin{lemma}
\label{equiv_constraints}
The inequalities (\ref{deterministic_contraction}) and (\ref{Mcon}) are equivalent to
\begin{align}
\label{deterministic_contraction_tilde}
&\dot{\tilde{W}}(x,t)-\sym{\left(\frac{\partial f(x,t)}{\partial x}\tilde{W}(x,t)\right)} \succeq 2\alpha \tilde{W}(x,t),~\forall x,t \\
\label{W_tilde}
&I \preceq \tilde{W}(x,t) \preceq \chi I,~\forall x,t
\end{align}
respectively, where $\chi = \overline{\omega}/\underline{\omega}$, $\tilde{W} = \nu W$, and $\nu = 1/\underline{\omega}$.
\end{lemma}
\begin{IEEEproof}
Since $\nu = 1/\underline{\omega} > 0$ and $W(x,t) \succ 0$, multiplying (\ref{deterministic_contraction}) by $\nu$ and then by $W(x,t)$ from both sides preserves matrix definiteness and the resultant inequalities are equivalent to the original ones~\cite[pp. 114]{lmi}. These operations yield (\ref{deterministic_contraction_tilde}). Next, since $M(x,t) \succ 0$, there exists a unique $\mu(x,t) \succ 0$ \st{} $M = \mu^2$. Multiplying (\ref{Mcon}) by $\mu^{-1}$ from both sides gives $\underline{\omega}I \preceq W(x,t) \preceq \overline{\omega}I$ as we have $\underline{\omega},\overline{\omega} > 0$ and $(\mu^{-1})^2 = W$. We get (\ref{W_tilde}) by multiplying this inequality by $\nu = 1/\underline{\omega}$.
\end{IEEEproof}
We are now ready to state and prove our main result on the convex optimization-based sampling.
\begin{theorem}
\label{conv_equiv_thm}
Consider the convex optimization problem:
\begin{align}
    \label{convex_opt_ccm}
    &{J}_{CV}^* = \min_{\chi \in \mathbb{R},\tilde{W}\succ 0} \frac{\overline{d}\chi}{\alpha} \text{~~\st{} (\ref{deterministic_contraction_tilde}) and (\ref{W_tilde})}
\end{align}
where $\chi$ and $\tilde{W}$ are defined in Lemma~\ref{equiv_constraints}, and $\alpha>0$ and $\overline{d}=\sup_{t \geq 0}\|d(t)\|$ are assumed to be given. Then ${J}_{NL}^*={J}_{CV}^*$.
\end{theorem}
\begin{IEEEproof}
By definition, we have ${\overline{d}\overline{\omega}}/{(\alpha\underline{\omega})}={\overline{d}\chi}/{\alpha}$.
Since (\ref{deterministic_contraction}) and (\ref{Mcon}) are equivalent to (\ref{deterministic_contraction_tilde}) and (\ref{W_tilde}) by Lemma~\ref{equiv_constraints}, rewriting the objective in the original problem (\ref{nonlin_opt_ccm}) using this equality completes the proof.
\end{IEEEproof}
\begin{remark}
\label{howtochosealpha}
Since (\ref{deterministic_contraction_tilde}) and (\ref{W_tilde}) are independent of $\nu=1/\underline{\omega}$, the choice of $\nu$ does not affect the optimal value of the minimization problem in Theorem~\ref{conv_equiv_thm}. In practice, as we have $\sup_{x,t}\|M(x,t)\|\leq1/\underline{\omega}=\nu$ by (\ref{Mcon}), it can be used as a penalty to optimally adjust the induced 2-norm of estimation and control gains when the problem explicitly depends on $\nu$ (see Sec.~\ref{example} for details). Also, although $\alpha$ is fixed in Theorem~\ref{conv_equiv_thm}, it can be found by a line search as will be demonstrated in Sec. \ref{simulation}. 
\end{remark}
\begin{remark}
The problem (\ref{convex_opt_ccm}) can be solved as a finite-dimensional problem by using backward difference approximation, $\dot{\tilde{W}}(x(t_i),t_i) \simeq {({\tilde{W}}(x(t_i),t_i)-{\tilde{W}}(x(t_{i-1}),t_{i-1}))}/{\Delta t_i}$, where $\Delta t = t_i-t_{i-1}, \forall i$ with $\Delta t \gg \Delta t^2 > 0$, and by discretizing it along a pre-computed system trajectory $\{x(t_i)\}_{i=0}^N$.
\end{remark}
\subsection{Deep LSTM-RNN Training}
Instead of directly using sequential data of optimal contraction metrics $\{M(x(t_i),t_i)\}^{N}_{i=0}$ for neural network training, the positive definiteness of $M(x,t)$ is utilized to reduce the dimension of the target output $\{y_i\}_{i=0}^N$ defined in Sec. \ref{preliminaries}.
\begin{lemma}
\label{cholesky}
A matrix $A \succ 0$ has a unique Cholesky decomposition, \ie{}, there exists a unique upper triangular matrix $U\in\mathbb{R}^{n\times n}$ with strictly positive diagonal entries \st{} $A = U^TU$.
\end{lemma}
\begin{IEEEproof}
See~\cite[pp. 441]{10.5555/2422911}.
\end{IEEEproof}
As a result of Lemma~\ref{cholesky}, we can select $\Theta (x,t)$ defined in Theorem~\ref{RobustCont} as the unique Cholesky decomposition of $M(x,t)$ and train the deep LSTM-RNN using only the non-zero entries of the unique upper triangular matrices $\{\Theta(x(t_i),t_i)\}^{N}_{i=0}$. We denote these nonzero entries as $\theta (x,t) \in \mathbb{R}^{\frac{1}{2}n(n+1)}$.
As a result, the dimension of the target data $\theta (x,t)$ is reduced by $n(n-1)/2$ without losing any information on $M(x,t)$.

The pseudocode to obtain an NCM depicted in Fig. \ref{ncmdrawing} is presented in Algorithm~\ref{NCMalg}.
The deep LSTM-RNN in Sec. \ref{preliminaries} is trained with the sequential state data $\{x(t_i)\}_{i=0}^N$ and the target data $\{\theta (x(t_i),t_i)\}^{N}_{i=0}$ using Stochastic Gradient Descent (SGD). We note these pairs will be sampled for multiple trajectories to increase sample size and to avoid overfitting. 
\begin{algorithm}
\SetKwInOut{Input}{Inputs}\SetKwInOut{Output}{Outputs}
\Input{Initial and terminal states $\{x_{s}(t_0),x_{s}(t_N)\}^{S}_{s=1}$}
\Output{NCM and steady-state bound $J_{CV}^*$ in  (\ref{convex_opt_ccm})}
\BlankLine
\textit{A. Sampling of Optimal Contraction Metrics} \\
\For{$s\leftarrow 1$ \KwTo $S$}{
Generate a trajectory $\{x_{s}(t_i)\}^{N}_{i=0}$ using $x_{s}(t_0)$ (could use $x_{s}(t_N)$ for  motion planning problems) \\
\For{$\alpha \in A_{\mathrm{linesearch}}$}{
Find $J_{CV}^*(\alpha,x_{s})$ and $\{\theta(x_{s}(t_i),t_i)\}^{N}_{i=0}$ by Th.~\ref{conv_equiv_thm}
}
Find $\alpha^*(x_{s}) = \text{arg}\min_{\alpha \in A_{\mathrm{linesearch}}}J_{CV}^*(\alpha,x_{s})$ \\
Save $\{\theta(x_{s}(t_i),t_i)\}^{N}_{i=0}$ for $\alpha = \alpha^*(x_{s})$
}
Obtain $J_{CV}^* = \max_{s}J_{CV}^*(\alpha^*(x_{s}),x_{s})$
\BlankLine
\textit{B. Deep LSTM-RNN Training} \\
Split data into a train set $\mathcal{S}_{\mathrm{train}}$ and test set $\mathcal{S}_{\mathrm{test}}$ \\
\For{$\mathrm{epoch} \leftarrow 1$ \KwTo $N_{\mathrm{epochs}}$}{
\For{$s \in \mathcal{S}_{\mathrm{train}}$}{Train the deep LSTM-RNN with $\{x_{s}(t_i)\}^{N}_{i=0}$, $\{\theta(x_{s}(t_i),t_i)\}^{N}_{i=0}$ using SGD}
Compute the test error for data in $\mathcal{S}_{\mathrm{test}}$ \\
\If{$\mathrm{test~error} < \epsilon$}{
\textbf{break}
}
}
\caption{NCM Algorithm}
\label{NCMalg}
\end{algorithm}
\section{NCM-Based Optimal Estimation and Control}
\label{example}
This section delineates how to construct an NCM offline and utilize it online for state estimation and feedback control.
\subsection{Problem Statement}
We apply an NCM to the state estimation problem for the following nonlinear system with bounded disturbances:
\begin{align}
    \label{orig_dynamics}
    \dot{x} = f(x,t)+B(x,t)d_1(t),~y(t) = h(x,t)+G(x,t)d_2(t)
\end{align}
where $d_1: \mathbb{R}_{\geq 0} \to \mathbb{R}^{k_1}$, $B:\mathbb{R}^n\times\mathbb{R}_{\geq 0} \to \mathbb{R}^{n\times k_1}$, $y: \mathbb{R}_{\geq 0} \to \mathbb{R}^m$, $d_2: \mathbb{R}_{\geq 0} \to \mathbb{R}^{k_2}$, $h:\mathbb{R}^n\times\mathbb{R}_{\geq 0} \to \mathbb{R}^{m}$, and $G:\mathbb{R}^n\times\mathbb{R}_{\geq 0} \to \mathbb{R}^{m\times k_2}$ with  $\overline{d}_1 = \sup_t\|d_1(t)\| < +\infty$ and $\overline{d}_2 = \sup_t\|d_2(t)\|  < +\infty$. Let $W = M(\hat{x},t)^{-1} \succ 0$, $A(x,t) = (\partial f/\partial x)$, and $C(x,t) = (\partial h/\partial x)$. We design an estimator as 
\begin{align}
    \label{est_dynamics}
    &\dot{\hat{x}} = f(\hat{x},t)+M(\hat{x},t)C(\hat{x},t)^T(y-h(\hat{x},t))\\
    \label{ekf_con}
    &\dot{W}+W A(\hat{x},t)+A(\hat{x},t)^TW-2C(\hat{x},t)^TC(\hat{x},t) \preceq -2\alpha W
\end{align}
where $\alpha > 0$. The virtual system of (\ref{orig_dynamics}) and (\ref{est_dynamics})  is given as
\begin{align}
    \label{detvd}
    \dot{q} =& f(q,t)+M(\hat{x},t)C(\hat{x},t)^T(h(x,t)-h(q,t))+d_e(q,t) 
\end{align}
where $d_e(q,t)$ is defined as $d_e(x,t) = B(x,t)d_1(t)$ and $d_e(\hat{x},t) = M(\hat{x},t)C(\hat{x},t)^TG(x,t)d_2(t)$. Note that (\ref{detvd}) has $q=x$ and $q=\hat{x}$ as its particular solutions. The differential dynamics of (\ref{detvd}) with $d_e = 0$ is given as
\begin{align}
    \label{detvd_dynamics}
    \delta \dot{q} =& (A(q,t)-M(\hat{x},t)C(\hat{x},t)^TC(q,t))\delta q.
\end{align}
\subsection{Nonlinear Stability Analysis}
We have the following lemma for deriving a condition to guarantee the local contraction of (\ref{detvd}) in Theorem~\ref{est_stability}.
\begin{lemma}
    \label{neighbor_lemma}
    If (\ref{ekf_con}) holds for $t\geq0$, there exists $r(t) > 0$ \st{}
    \begin{align}
    \label{ekf_virtual}
    2\gamma W+\dot{W}+\sym{}(W A(q,t))-\sym(C(\hat{x},t)^TC(q,t)) \preceq 0
    \end{align}
    for all $q(t)$ with $\|q(t)-\hat{x}(t)\| \leq r(t)$, where $0 < \gamma < \alpha$. 
\end{lemma}
\begin{IEEEproof}
See~Lemma 2 of~\cite{6849943} or Theorem 1 of~\cite{observer}.
\end{IEEEproof}
The following theorem along with this lemma guarantees the exponential stability of the estimator (\ref{est_dynamics}).
\begin{theorem}
\label{est_stability}
Suppose that there exist positive constants $\underline{\omega}$, $\overline{\omega}$, $\overline{b}$, $\bar{c}$, $\bar{g}$, and $\rho$ \st{} $\underline{\omega}I \preceq W(\hat{x},t) \preceq \overline{\omega}I$, $\|B(x,t)\| \leq \overline{b}$, $\|C(\hat{x},t)\| \leq \bar{c}$, $\|G(x,t)\| \leq \bar{g}$, and $r(t) \geq \rho,~\forall \hat{x},x,t$, where $r(t)$ is defined in Lemma~\ref{neighbor_lemma}. If (\ref{ekf_con}) holds and $R_e(0)+\overline{D}_e/{\gamma} \leq \sqrt{\underline{\omega}} \rho$, where $R_e(t) = \int_{\hat{x}}^x\|\Theta(\hat{x},t)\delta q(t)\|$ with $W = \Theta^T \Theta$ and $\overline{D}_e = \overline{d}_1\overline{b}\sqrt{\overline{\omega}}+{\overline{d}_2\bar{c}\bar{g}}/{\sqrt{\underline{\omega}}}$, then the distance between the trajectory of (\ref{orig_dynamics}) and (\ref{est_dynamics}) is exponentially bounded as follows:
\begin{align}
    \label{est_ss}
    \int_{\hat{x}}^x\|\delta q\| \leq \frac{R_e(0)}{\sqrt{\underline{\omega}}}e^{-\gamma t}+ \frac{\overline{d}_1\overline{b}}{\gamma}\chi+\frac{{\overline{d}_2\bar{c}\bar{g}}}{\gamma}\nu
\end{align}
where $\chi = {\overline{\omega}}/{\underline{\omega}}$, $\nu = {1}/{\underline{\omega}}$, and $0 < \gamma < \alpha$.
\end{theorem}
\begin{IEEEproof}
Using (\ref{detvd_dynamics}), we have $d(\|\Theta \delta q\|^2)/dt = \delta q^T(\dot{W}+\sym(W A(q,t))-\sym(C(\hat{x},t)^TC(q,t)))\delta q$ when $d_e = 0$. This along with (\ref{ekf_virtual}) gives $\dot{R}_e(t) \leq -{\gamma} R_e(t)$ in the region where Lemma~\ref{neighbor_lemma} holds. Thus, using the bound $\|\Theta(\hat{x}(t),t)d_e(q,t)\| \leq \overline{D}_e$, we have $\sqrt{\underline{\omega}}\int_{\hat{x}}^{x}\|\delta q\| \leq {R_e(0)}e^{-\gamma t}+{\overline{D}_e}/\gamma$ by the same proof as for Theorem~\ref{RobustCont}. Rewriting this with $\chi$, $\nu$, and $1 \leq \sqrt{\chi} \leq \chi$ yields (\ref{est_ss}). This also implies that $\sqrt{\underline{\omega}}\|x-\hat{x}\| \leq R_e(0)+\overline{D}_e/{\gamma},~\forall t$. Hence, the sufficient condition for $\|q-\hat{x}\|$ in Lemma~\ref{neighbor_lemma} reduces to the one required in this theorem.
\end{IEEEproof}
\subsection{Convex Optimization-based Sampling (CV-STEM)}
We have the following proposition to sample optimal contraction metrics for the NCM-based state estimation.
\begin{proposition}
\label{est_prop}
$M(\hat{x},t)$ that minimizes an upper bound of $\lim_{t\to\infty} \int_{\hat{x}}^x\|\delta q\|$ is found by the convex optimization problem:
\begin{align}
    \label{convex_opt_estimator}
    &{J}_{CVe}^* = \min_{\nu>0,\chi \in \mathbb{R},\tilde{W} \succ 0} \frac{\overline{d}_1\overline{b}}{\gamma}\chi+\frac{{\overline{d}_2\bar{c}\bar{g}}}{\gamma}\nu\\
    &\text{\st{}}\text{  $\dot{\tilde{W}}+\tilde{W}A+A^T\tilde{W}-2\nu C^TC \preceq -{2\alpha} \tilde{W}$ and $I \preceq \tilde{W} \preceq \chi I$} \nonumber
\end{align}
where $\chi = {\overline{\omega}}/{\underline{\omega}}$, $\nu = {1}/{\underline{\omega}}$, $\tilde{W} = \nu W$, and $0 < \gamma < \alpha$. The arguments of $A(\hat{x},t)$, $C(\hat{x},t)$, and $\tilde{W}(\hat{x},t)$ are omitted for notational simplicity.
\end{proposition}
\begin{IEEEproof}
Multiplying (\ref{ekf_con}) and $\underline{\omega}I \preceq W(\hat{x},t) \preceq \overline{\omega}I,~\forall \hat{x},t$ by $\nu$ yields the constraints of (\ref{convex_opt_estimator}). Then Theorem~\ref{conv_equiv_thm} with the objective function given in (\ref{est_ss}) of Theorem~\ref{est_stability} as $t \to \infty$ yields (\ref{convex_opt_estimator}).
\end{IEEEproof}

We have an analogous result for state feedback control.
\begin{corollary}
\label{con_cor}
Consider the following system and a state feedback controller $u(t)$ with the bounded disturbance $d(t)$:
\begin{align}
    \label{control_dynamics}
    &\dot{x} = f(x,t)+B_1(x,t)u+B_2(x,t)d(t) \\
    \label{con_riccati}
    &\dot{W}-A(x,t)W-W A(x,t)^T+2B_1(x,t)B_1(x,t)^T \succeq 2\alpha W
\end{align}
where $u = -B_1(x,t)^TM(x,t)x$, $B_1:\mathbb{R}^n\times\mathbb{R}_{\geq0}\to\mathbb{R}^{n\times m}$, $B_2:\mathbb{R}^n\times\mathbb{R}_{\geq0}\to\mathbb{R}^{n\times k}$, $W=M^{-1}\succ 0$, $\alpha>0$, and $A$ is a matrix defined as $A(x,t)x = f(x,t)$, assuming that $f(x,t) = 0$ at $x = 0$ \cite{sddre,CIMEN20083761}. Suppose there exist positive constants $\underline{\omega}$, $\overline{\omega}$, and $\overline{b}_2$ \st{} $\underline{\omega}I \preceq W(x,t) \preceq \overline{\omega}I$ and $\|B_2(x,t)\| \leq \overline{b}_2,~\forall x,t$.
Then $M(x,t)$ that minimizes an upper bound of $\lim_{t\to\infty} \int_{0}^x\|\delta q\|$ can be found by the following convex optimization problem:
\begin{align}
    \label{convex_opt_controller}
    &{J}_{CVc}^* = \min_{\nu > 0,\chi \in \mathbb{R},\tilde{W} \succ 0} \frac{\overline{b}_2\overline{d}}{\alpha}\chi+\lambda\nu\\
    &\text{\st{}}\text{  $-\dot{\tilde{W}}+A\tilde{W}+\tilde{W}A^T-2\nu B_1B_1^T \preceq -{2\alpha}\tilde{W}$ and $I \preceq \tilde{W} \preceq \chi I$} \nonumber
\end{align}
where $\chi = {\overline{\omega}}/{\underline{\omega}}$, $\nu = {1}/{\underline{\omega}}$, $\tilde{W} = \nu W$, and $\lambda>0$ is a user-defined constant. The arguments of $A(x,t)$, $B_1(x,t)$, and $\tilde{W}(x,t)$ are omitted for notational simplicity.
\end{corollary}
\begin{IEEEproof}
The system with $q=x,0$ as its particular solutions is given by $\dot{q} = (A(x,t)-B_1(x,t)B_1(x,t)^TM(x,t))q+d_c(q,t)$, where $d_c(x,t) = B_2(x,t)d(t)$ and $d_c(0,t) = 0$. Since we have $\|d_c(q,t)\| \leq \overline{b}_2\overline{d}$ and the differential dynamics is 
\begin{align}
\label{differential_con}
    \delta \dot{q} = (A(x,t)-B_1(x,t)B_1(x,t)^TM(x,t))\delta q
\end{align}
when $d_c = 0$, we get $\lim_{t\to \infty}\int_{0}^x\|\delta q\| \leq \overline{b}_2\overline{d}\chi/\alpha$ by the same proof as for Theorem~\ref{est_stability} with (\ref{detvd_dynamics}) replaced by (\ref{differential_con}), (\ref{ekf_virtual}) by (\ref{con_riccati}), and $R_e(t)$ by $R_c(t)=\int_{0}^x\|\Theta(x,t )\delta q(t)\|$, where $M = \Theta^T\Theta$. (\ref{convex_opt_controller}) then follows as in the proof of Proposition~\ref{est_prop}, where $\lambda \geq 0$ is for penalizing excessively large control inputs through $\nu \geq \sup_{x,t}\|M(x,t)\|$ (see Remark~\ref{howtochosealpha}).
\end{IEEEproof}
\subsection{NCM Construction and Interpretation} Algorithm~\ref{NCMalg} along with Proposition~\ref{est_prop} and Corollary~\ref{con_cor} returns NCMs to compute $\hat{x}(t)$ of (\ref{est_dynamics}) and $u(t)$ of (\ref{control_dynamics}) for state estimation and control in real-time. They also provide the bounded error tube (see Theorem~\ref{RobustCont}, \cite{chung2013phase,7989693}) for robust motion planning problems as will be seen in Sec.~\ref{simulation}.

The similarity of Corollary~\ref{con_cor} to Proposition~\ref{est_prop} stems from the estimation and control duality due to the differential nature of contraction analysis as is evident from (\ref{detvd_dynamics}) and (\ref{differential_con}). Analogously to the discussion of the Kalman filter and LQR duality in LTV systems, this leads to two different interpretations on the weight of $\nu$ ($\overline{d}_2\bar{c}\bar{g}/\gamma$ in (\ref{convex_opt_estimator}) and $\lambda$ in (\ref{convex_opt_controller})). As discussed in Remark~\ref{howtochosealpha}, one way is to see it as a penalty on the induced 2-norm of feedback gains. Since $\overline{d}_2=0$ in (\ref{convex_opt_estimator}) means no noise acts on $y(t)$, it can also be viewed as an indicator of how much we trust the measurement $y(t)$: the larger the weight of $\nu$, the less confident we are in $y(t)$. These agree with our intuition as smaller feedback gains are suitable for systems with larger measurement uncertainty.
\section{Simulation}
\label{simulation}
\renewcommand\UrlFont{\rm}
The NCM framework is demonstrated using Lorentz oscillator state estimation and spacecraft motion planning and control problems. CVXPY~\cite{cvxpy} with the MOSEK solver~\cite{mosek} is used to solve convex optimization problems. A Python implementation is available at {\color{caltechgreen}\underline{\url{https://github.com/astrohiro/ncm}}}.
\subsection{State Estimation of Lorenz Oscillator}
We first consider state estimation of the Lorentz oscillator with bounded disturbances described as $\dot{x}=f(x)+d_1(t)$ and $y = Cx+d_2(t)$, where $f(x) =[\sigma(x_2-x_1),x_1(\rho-x_3)-x_2,x_1 x_2-\beta x_3]^T$, $x = [x_1,x_2,x_3]^T$, $\sigma = 10$, $\beta = 8/3$, $\rho = 28$, $C = [1~0~0]$, $\sup_t\|d_1(t)\| = \sqrt{3}$, and $\sup_t\|d_2(t)\| = 1$. We use $dt = 0.1$ for integration, with one measurement $y$ per $dt$.
\subsubsection{Sampling of Optimal Contraction Metrics}
\begin{figure}
    \centering
    \includegraphics[width=70mm]{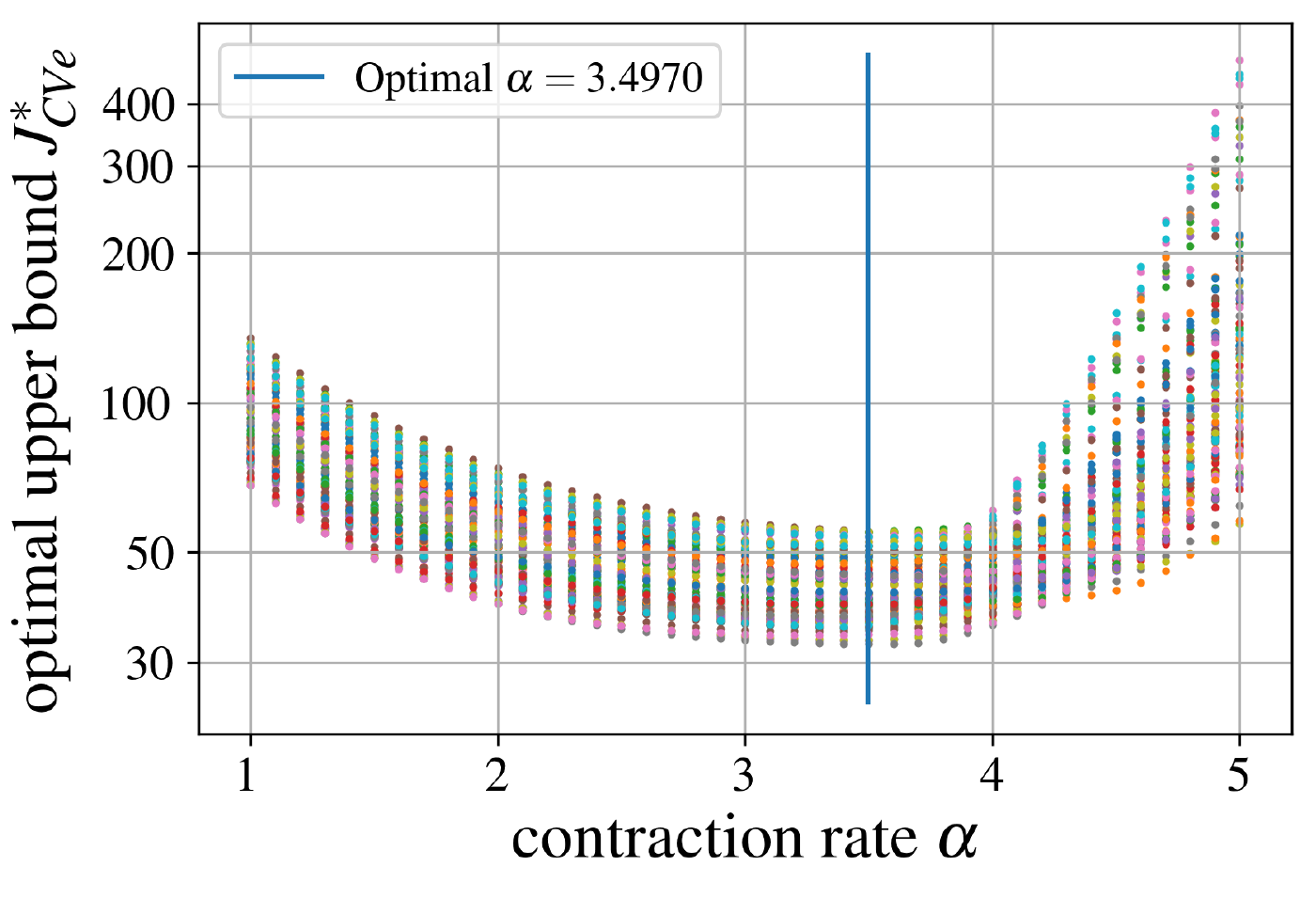}
    \caption{Upper bound of steady-state errors as a function of $\alpha$: each curve with a different color corresponds to a trajectory with a different initial condition.}
    \label{lorenz_micro}
\end{figure}

Using Proposition~\ref{est_prop}, we sample the optimal contraction metric along $100$ trajectories with uniformly distributed initial conditions ($-10 \leq x_i \leq 10,~i = 1,2,3$). Figure~\ref{lorenz_micro} plots ${J}_{CVe}^*$ in (\ref{convex_opt_estimator}) for $100$ different trajectories and the optimal $\alpha$ is found to be $\alpha = 3.4970$. The optimal estimator parameters averaged over $100$ trajectories for $\alpha = 3.4970$ are summarized in Table~\ref{lorenz_optvals}.
\begin{table}
\caption{NCM Estimator Parameters for $\alpha = 3.4970$ with ${J}_{CVe}^*$ and MSE Averaged over $100$ Trajectories}
\label{lorenz_optvals}
\centering
\begin{tabular}{|c|c|c|c|c|} 
 \hline
 parameters & $\nu$ & $\chi$ & ${J}_{CVe}^*$ & MSE \\
 \hline
 values & $1.3375 \times 10^2$ & $9.2977$ & $42.852$ & $1.0190 \times 10^{-3}$ \rule[0mm]{0mm}{2.7mm}\\
 \hline
\end{tabular}
\end{table}
\subsubsection{Deep LSTM-RNN Training}
\begin{figure}
    \centering
    \includegraphics[width=87mm]{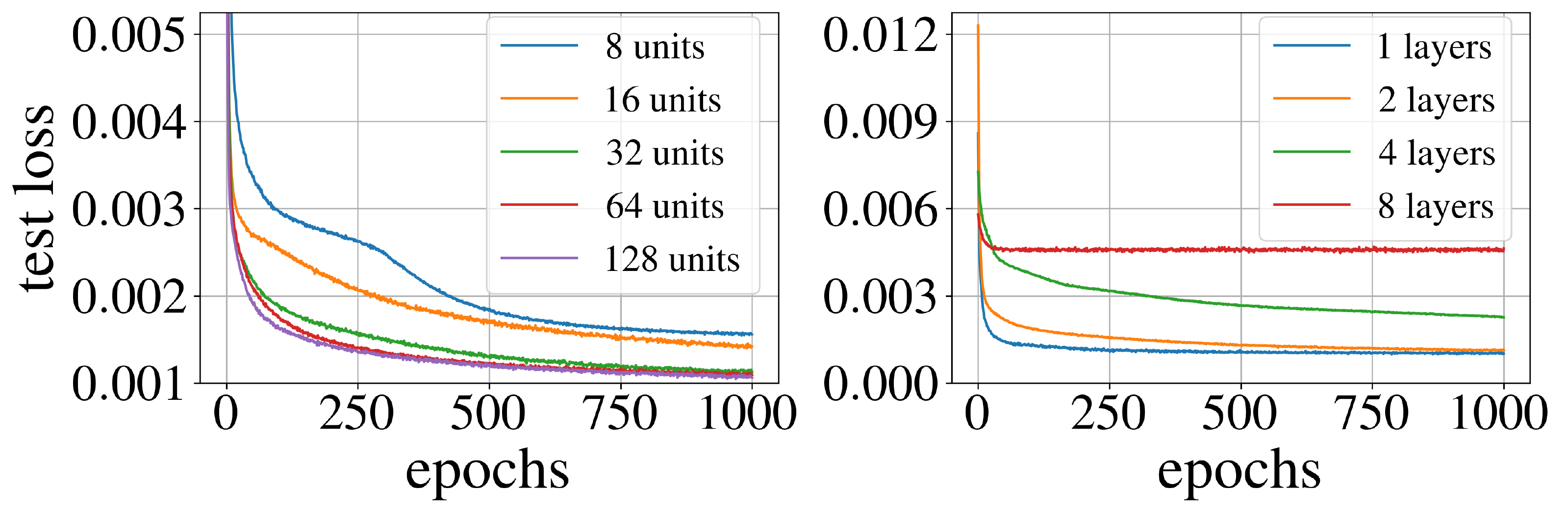}
    \caption{LSTM-RNN test loss with 2 layers (left) and 32 hidden units (right).}
    \label{model_selection}
\end{figure}

A deep LSTM-RNN is trained using Algorithm~\ref{NCMalg} and Proposition~\ref{est_prop} with the sequential data $\{\{(x_{s}(t_i),\theta_{s}(x(t_i))\}_{i=0}^N\}_{s=1}^S$ sampled over the $100$ different trajectories ($S = 100$). Note that $\theta_{s}(x(t_i))$ are standardized and normalized to make the SGD-based learning process stable. Figure~\ref{model_selection} shows the test loss of the LSTM-RNN models with different number of layers and hidden units. We can see that the models with more than $2$ layers overfit and those with less than $32$ hidden units underfit to the training samples. Thus, the number of layers and hidden units are selected as $2$ and $64$, respectively. The resultant MSE of the trained LSTM-RNN is shown in Table~\ref{lorenz_optvals}.
\subsubsection{State Estimation with an NCM}
The estimation problem is solved using the NCM, sampling-based CV-STEM~\cite{mypaper,mypaperTAC}, and Extended Kalman Filter (EKF) with $\sup_t\|d_1(t)\| = 20\sqrt{3}$, and $\sup_t\|d_2(t)\| = 20$. We use $x(0) = [-1.0,2.0,3.0]^T$ and $\hat{x}(0) = [150.1,-1.5,-6]^T$ for the actual and estimated initial conditions, respectively. The EKF weight matrices are selected as $R = 20I$ and $Q = 10I$.

\begin{figure}
    \centering
    \includegraphics[width=70mm]{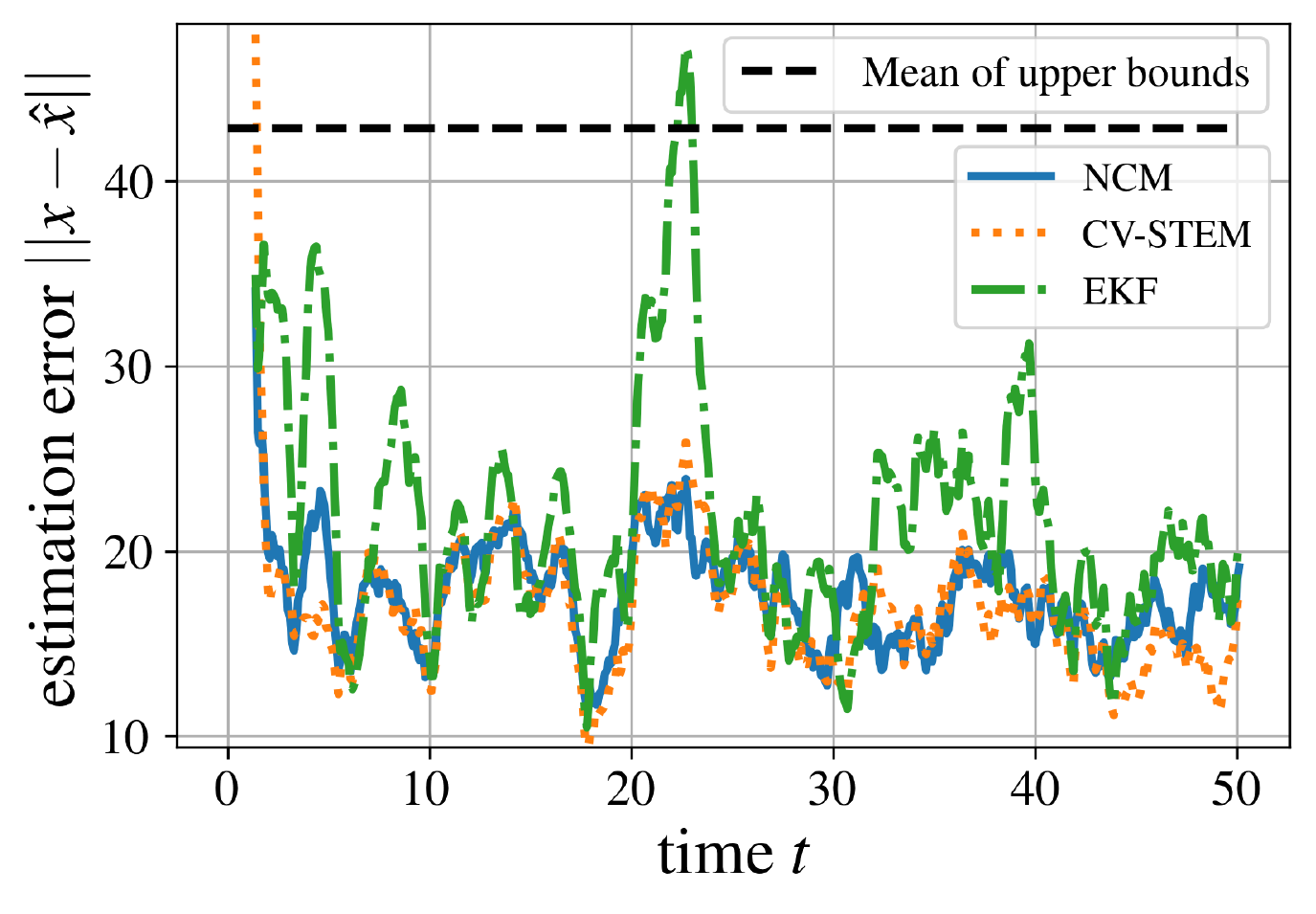}
    \caption{Lorentz oscillator state estimation error smoothed using a $15$-point moving average filter.}
    \label{lorenz_plot}
\end{figure}

Figure~\ref{lorenz_plot} shows the smoothed estimation error $\|x(t)-\hat{x}(t)\|$ using a $15$-point moving average filter. The errors of the NCM and CV-STEM estimators are below the optimal upper bound while the EKF has a larger error compared to the other two. 
As expected from the small MSE of Table~\ref{lorenz_optvals}, the estimation error of the NCM estimator shows a trend similar to that of the sampling-based CV-STEM estimator without losing its estimation performance.
\subsection{Spacecraft Optimal Motion Planning}
We consider an optimal motion planning problem of the planar spacecraft dynamical system, given as $\dot{x} = Ax+B(x)u+d(t)$, where $u \in \mathbb{R}^8$, $\sup_t\|d(t)\| = 0.15$, and $x = [p_x,p_y,\phi,\dot{p}_x,\dot{p}_y,\dot{\phi}]^T$ with $p_x$, $p_y$, and $\phi$ being the horizontal coordinate, vertical coordinate, and yaw angle of the spacecraft, respectively. The constant matrix $A$ and the state-dependent actuation matrix $B(x)$ are defined in \cite{SCsimulator}. All the parameters of the spacecraft are normalized to $1$.
\subsubsection{Problem Formulation}
In the planar field, we have $6$ circular obstacles with radius $3$ located at $(p_x,p_y) = (0,11)$, $(5,3)$, $(8,11)$, $(13,3)$, $(16,11)$, and $(21,3)$. The goal of the motion planning problem is to find an optimal trajectory that avoids these obstacles and minimize $\int_0^{50}\|u(t)\|^2dt$ subject to input constraints $0 \leq u_i(t) \leq 1,\forall i,\forall t$ and the dynamics constraints. The initial and terminal condition are selected as $x(0) =[0,0,\pi/12,0,0,0]^T$ and $x(t_N) =[20,18,0,0,0,0]^T$. Following the same procedure described in the state estimation problem, the optimal control parameters and the MSE of the LSTM-RNN trained using Algorithm~\ref{NCMalg} with Corollary~\ref{con_cor} are determined as shown in Table~\ref{sc_optvals}. 
\begin{table}
\caption{NCM Control Parameters for $\alpha = 0.58$ with ${J}_{CVc}^*$ and MSE Averaged over $100$ Trajectories}
\label{sc_optvals}
\centering
\begin{tabular}{|c|c|c|c|c|} 
 \hline
 parameters & $\nu$ & $\chi$ & ${J}_{CVc}^*$ & MSE \\
 \hline
 values & $6.2090 \times 10^2$ & $3.0116$ & $8.9524$ & $3.4675\times 10^{-4}$ \rule[0mm]{0mm}{2.7mm}\\
 \hline
\end{tabular}
\end{table}
\subsubsection{Motion Planning with an NCM}
\begin{figure}
    \centering
    \includegraphics[width=70mm]{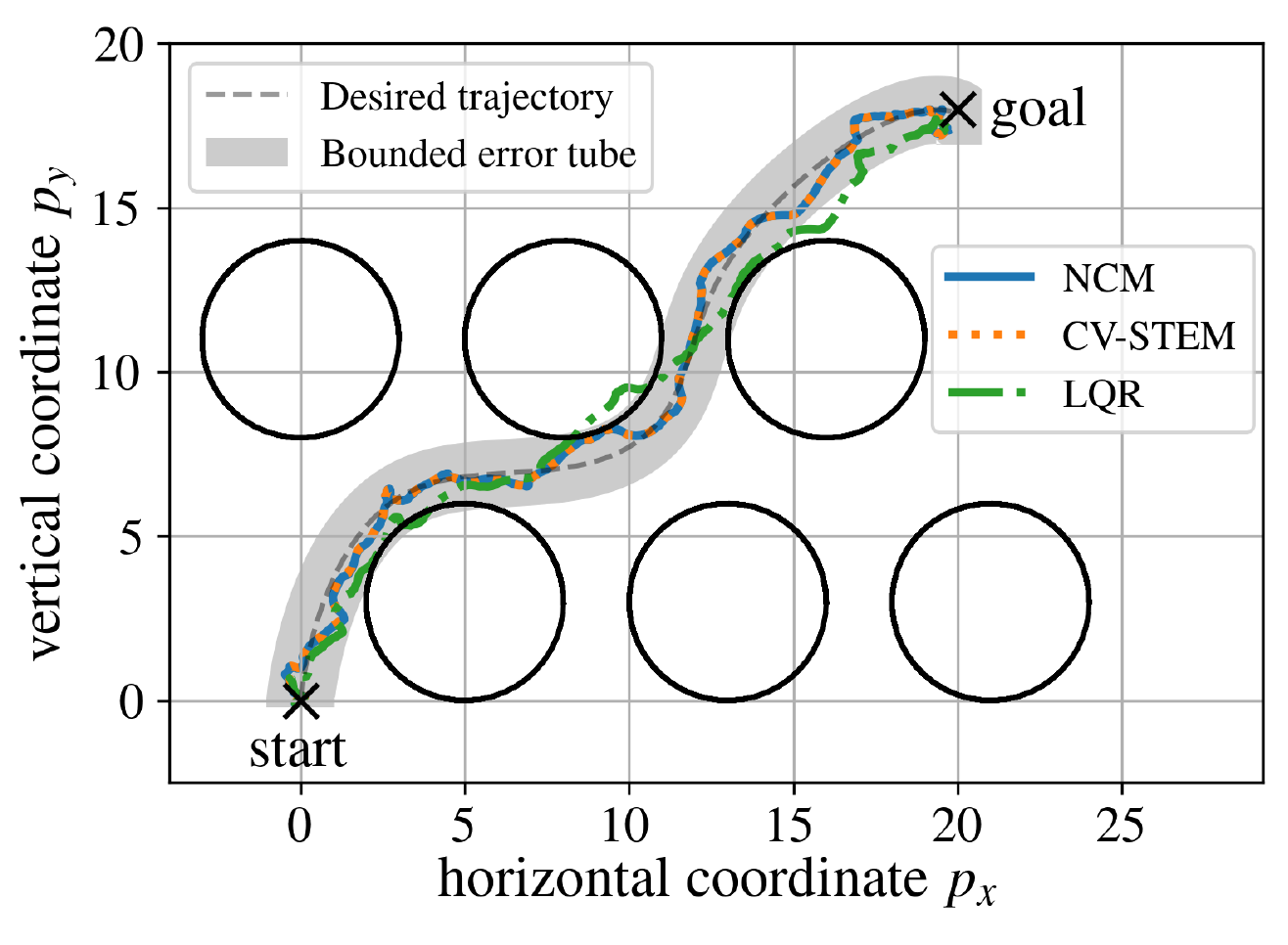}
    \caption{Spacecraft motion $(p_x,p_y)$ on a planar field.}
    \label{robust_SC}
\end{figure}

Given an NCM, we can solve a robust motion planning problem, where the state constraint is now described by the bounded error tube (see Theorem~\ref{RobustCont}) with radius $R_\mathrm{tube} = \overline{d}(\sqrt{\chi}/\alpha) = 0.4488$. Figure~\ref{robust_SC} shows the spacecraft motion $(p_x,p_y)$ on a planar field, computed using the NCM, sampling-based CV-STEM~\cite{mypaper,mypaperTAC}, and baseline LQR control with $Q = 2.4I$ and $R = I$ which does not account for the disturbance.
As summarized in Table~\ref{con_eff}, input constraints $0 \leq u_i(t) \leq 1,\forall i, \forall t$ are satisfied and the three controllers use a similar amount of control effort.
All the controllers except the LQR keep their trajectories within the tube avoiding collision with the circular obstacles, even under the presence of disturbances as depicted in Fig.~\ref{robust_SC}. Also, the NCM controller predicts the computationally-expensive CV-STEM controller with the small MSE as given in the last column of Table \ref{sc_optvals}.
\begin{table}
\caption{Control Performances for Spacecraft Motion Planning Problem}
\label{con_eff}
\centering
\begin{tabular}{|c|c|c|c|} 
 \hline
  & $\int_0^{50}\|u(t)\|^2dt$ & $\min_i(\inf_t u_i(t))$ & $\max_i(\sup_t u_i(t))$  \\
 \hline
 NCM & $4.8959 \times 10$ & $-1.2728 \times 10^{-8}$ & $1.0000$ \rule[0mm]{0mm}{2.7mm}\\
 \hline
 CV-STEM & $4.8019 \times 10$ & $-4.1575 \times 10^{-8}$ & $1.0000$ \rule[0mm]{0mm}{2.7mm}\\
 \hline
 LQR & $4.9260 \times 10$ & $0.0000$ & $1.0000$ \rule[0mm]{0mm}{2.7mm}\\
 \hline
\end{tabular}
\end{table}
\section{Conclusion}
\label{conclusion}
In this paper, we present a Neural Contraction Metric (NCM), a deep learning-based  global approximation of an optimal contraction metric for online nonlinear estimation and control.
The novelty of the NCM approach lies in that: 1) data points of the optimal contraction metric are sampled offline by solving a convex optimization problem, which minimizes an upper bound of the steady-state Euclidean distance between the perturbed and unperturbed trajectories without assuming any hypothesis function space, and 2) the deep LSTM-RNN is constructed to model the sampled metrics with arbitrary accuracy. The superiority of its performance is validated through numerical simulations on state estimation and optimal motion planning problems.
\bibliographystyle{IEEEtran}
\bibliography{ms}

\end{document}